# scGSDR: Harnessing Gene Semantics for Single-Cell Pharmacological Profiling

Yu-An Huang[1,2, #,*], Xiyue Cao[1,#], Zhu-Hong You[1,*], Yue-Chao Li[1], Xuequn Shang[1], Zhi-An Huang[3]

[1] School of Computer Science, Northwestern Polytechnical University

[2] Research & Development Institute of Northwestern Polytechnical University in Shenzhen

[3] City University of Hong Kong (Dongguan)

#equal contribution

*corresponding authors

**Abstract**

The rise of single-cell sequencing technologies has revolutionized the exploration of drug resistance, revealing the crucial role of cellular heterogeneity in advancing precision medicine. By building computational models from existing single-cell drug response data, we can rapidly annotate cellular responses to drugs in subsequent trials. To this end, we developed scGSDR, a model that integrates two computational pipelines grounded in the knowledge of cellular states and gene signaling pathways, both essential for understanding biological gene semantics. scGSDR enhances predictive performance by incorporating gene semantics and employs an interpretability module to identify key pathways contributing to drug resistance phenotypes. Our extensive validation, which included 16 experiments covering 11 drugs, demonstrates scGSDR's superior predictive accuracy, when trained with either bulk-seq or scRNA-seq data, achieving high AUROC, AUPR, and F1 Scores. The model's application has extended from single-drug predictions to scenarios involving drug combinations. Leveraging pathways of known drug target genes, we found that scGSDR's cell-pathway attention scores are biologically interpretable, which helped us identify other potential drug-related genes.

Literature review of top-ranking genes in our predictions such as BCL2, CCND1, the AKT family, and PIK3CA for PLX4720; and ICAM1, VCAM1, NFKB1, NFKBIA, and RAC1 for Paclitaxel confirmed their relevance. In conclusion, scGSDR, by incorporating gene semantics, enhances predictive modeling of cellular responses to diverse drugs, proving invaluable for scenarios involving both single drug and combination therapies and effectively identifying key resistance-related pathways, thus advancing precision medicine and targeted therapy development.

## 1. Background

Predicting drug responses based on individual transcriptomic profiles serves as a cornerstone for precision medicine, enabling tailored therapeutic strategies that optimize efficacy and minimize the risk of treatment failure for each patient. Extensive research consistently demonstrates that treatment failure is intricately linked to cellular heterogeneity, which fosters the emergence of rare subpopulations of malignant cells(1, 2). These cells exhibit distinct epigenetic and transcriptional states, rendering them resistant to a diverse array of anticancer drugs and thus preventing the total eradication of the tumor population(3). A thorough understanding of such a complex genetic and molecular landscape within a tumor is crucial for devising more effective, personalized therapy strategies that specifically target the unique characteristics of all cancerous subpopulations(4).

Single-cell sequencing technologies have emerged as pivotal methodologies in delineating tumor heterogeneity(5). These technologies facilitate a granular analysis of the diverse cellular environments within tumors, enabling a deeper understanding of the molecular underpinnings of various cancer subtypes(6). Recent efforts have expanded to incorporate the use of single-cell sequencing to elucidate the mechanisms underlying drug responses in cancer treatment(7). Although these studies are still in their preliminary stages and the dataset is relatively limited, they represent a critical advancement in understanding how individual cells within a tumor may respond to therapeutic interventions(8). Given the crucial significance of these findings, there has been a notable increase in the deployment of sophisticated computational models utilizing deep learning and machine learning to precisely annotate cellular drug response.

Current computational models used to predict cellular drug responses primarily employ databases such as the Cancer Cell Line Encyclopedia (CCLE)(9) and the Genomics of Drug Sensitivity in Cancer (GDSC)(10). These repositories provide large-scale expression profiles of cancer cell lines using bulk RNA sequencing (RNA-seq), which yields an averaged estimation of gene expression across various subtypes within tumor cell populations. However, the effective application of these models faces challenges due to the fundamental differences between bulk RNA-seq and single-cell RNA-seq (scRNA-seq) techniques. Bulk RNA-seq aggregates the gene expression data of thousands of cells, potentially masking the unique responses of individual cells crucial for precise drug response prediction. Therefore, the success of these computational models hinges on the effective transfer of knowledge from bulk RNA-seq data, used as the reference, to scRNA-seq data, used as the query. Two prime examples include SCAD(11), which adopts an adversarial learning approach by training a domain discriminator, and scDEAL(12), which utilizes a loss function based on Maximum Mean Discrepancy (MMD) that is minimized to retain similarity between bulk and single-cell data.

With the increasing availability of single-cell sequencing data in drug response testing, it has become feasible to train a model using single-cell data to predict the response of single cells from other experiments to the same drug based on their gene expression profiles. However, training models with scRNA-seq data faces challenges such as dropout events that cause sparse gene expression and the limited number of genes related to drug mechanisms. This suggests that relying solely on domain adaptation for the knowledge transfer between datasets may not be sufficient for robust predictive modeling. Existing computational methods map all genes to the same learned feature space based on expression levels, thereby overlooking the distinct biological semantics attached to each gene. The biological semantics of genes, which reflect their functions and roles within the cell, are crucial for understanding their contribution to drug response mechanisms. Such semantics can be revealed through gene signaling pathways, which illustrate how genes execute specific biological functions via signal transduction. Such semantics, which can be revealed through gene signaling pathways that illustrate how genes execute specific biological functions via signal transduction, are also reflected in the macroscopic manifestations of cellular states, encompassing the cell's phenotype and behavior.

In this work, we introduce the scGSDR model (Single-cell Gene Semantics for Drug Response prediction), which employs a dual computational pipeline to integrate the prior knowledge of cellular states and gene signaling pathways, enhancing the prediction of drug responses at the single-cell level. The model was tested across three scenarios: using bulk-seq data as a reference, scRNA-seq data for single-drug predictions, and scRNA-seq for combination drug experiments. These applications demonstrate scGSDR's ability to handle both traditional and novel predictive challenges from various scenarios. We further show how the scGSDR model utilizes an attention mechanism interpretability module to discern the contribution of each pathway in each cell during the model inference process. It enables the identification of the most relevant gene pathways potentially leading to drug-resistant and drug-sensitive phenotypes within the dataset. Leveraging these pathways, we showcase a novel method for predicting drug action mechanisms and identifying genes associated with specific drugs. Specifically, we identified genes associated with the drug PLX4720, including BCL2, CCND1, the AKT family, and PIK3CA, as well as genes linked to the drug Paclitaxel, such as ICAM1, VCAM1, NFKB1, NFKBIA, and RAC1.

## 2. Results

### 2.1 Overview of scGSDR

We present the scGSDR (acronym for single-cell Gene Semantic Drug Response prediction) model depicted in Fig. 1 The proposed model considers two biological semantic perspectives of genes in scRNA-seq data through two computational pipelines, focusing on cellular states and cellular signaling pathways. In the first pipeline, tailored to assess cellular states, we utilize marker genes from 14 different cellular states as criteria for gene filtering. This process constructs cellular features, which are then mapped into an embedding space using a transformer module for cell representation. The second pipeline automatically learns attention matrices that define the association between each cell and various pathways, and constructs cell-cell graphs. These learned graphs, along with the gene expression profiles within the pathways, are input into a multi-graph fusion module, which maps them to another embedding space. Through both computational pipelines, each cell in the scRNA-seq single-cell matrix acquires two distinct embeddings, which are then integrated through feature fusion to produce the final fused embedding for annotating cellular drug responses.

The application of scGSDR model is designed to function in scenarios where scRNA-seq or cell line bulk-seq drug response data for a particular drug are used as a reference. Our model utilizes supervised learning to annotate drug responses in any scRNA-seq dataset for that drug. Considering that the reference data used during training inherently exhibit batch effects with the query data employed during testing, we incorporate a loss function augmented with domain adaptation (DA) learning to train our model, aiming to mitigate their discrepancies. Additionally, to address the potential issue of data imbalance between drug-resistant and drug-sensitive cells, the scGSDR model is optionally equipped with a loss function designed to correct distribution anomalies.

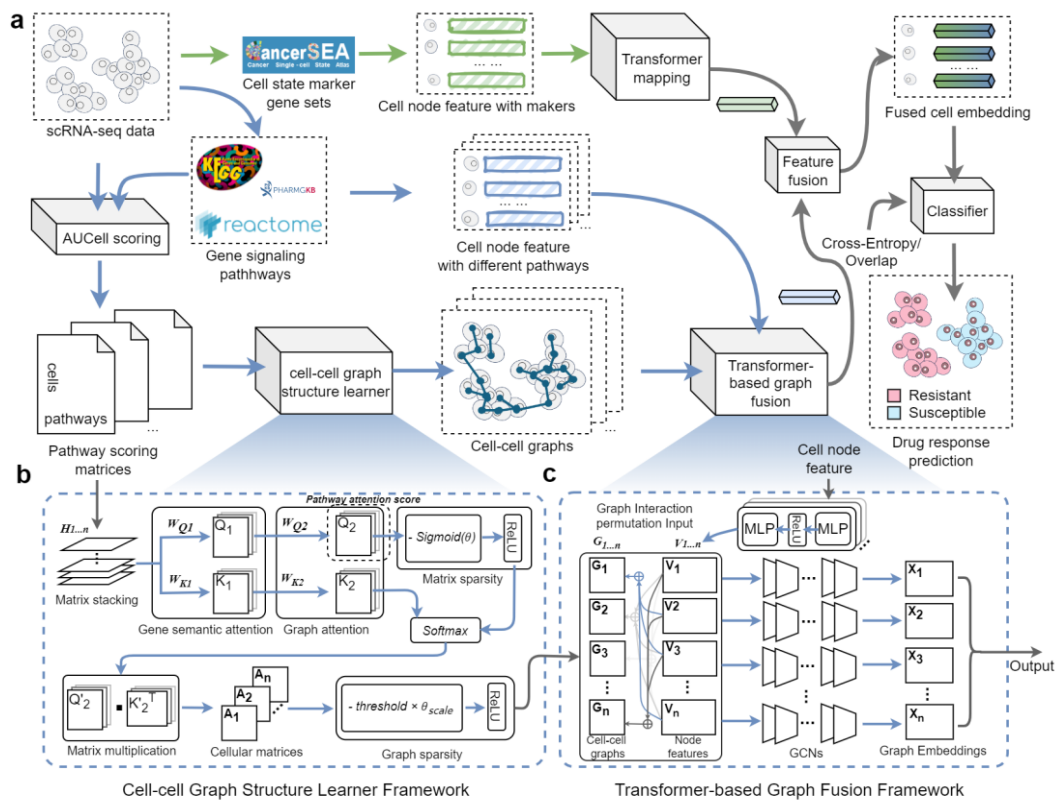

**Fig. 1** Overview of the scGSDR model. **a** The scGSDR model considers scRNA-seq data from two biological semantic perspectives of genes, utilizing two computational pipelines that employ cellular state information and cellular signaling pathways to construct cellular features. These features are subsequently integrated through feature fusion to produce the final embedding used for predicting cellular drug responses. **b** Detailed structure of the cell-cell graph structure learner. **c** Detailed structure of the transformer-based graph fusion component.

## 2.2 Evaluation of scGSDR on Bulk RNA-seq Reference Datasets

Since existing computational models for predicting drug responses in single-cell data primarily utilize cell line bulk-seq data, we sought to objectively compare their performance with that of our proposed scGSDR model. Therefore, we adopted the reference-query dataset pairs used in the SCAD model and employed the same five-fold cross-validation framework to evaluate the predictive performance of our model in training scenarios that utilize cell line bulk-seq data. In each fold, four-fifths of the query dataset along with their 'dataset labels' are used to train the DA classifier equipped in the scGSDR, with the remaining fifth reserved for testing. Importantly, as depicted in Fig. 3d, the term 'dataset labels' refers to identifiers that specify whether cells are derived from the reference or the query datasets, not to be confused with cell type labels. These dataset labels are utilized exclusively for training the DA classifier within the framework of five-fold cross-validation. Consequently, the cell type labels present in the query dataset do not contribute to the model training process.

In this section, we tested nine drugs (i.e., Afatinib, AR-42, Cetuximab, Etoposide, Gefitinib, NVP-TAE684, PLX4720, Sorafenib, and Vorinostat) across ten experiments. Notably, PLX4720 was evaluated in two separate experiments on the A375 and 451Lu cell lines. For each drug, the corresponding bulk-seq data set, sourced from the GDSC database, served as the reference for training the model, while the scRNA-seq data sets were used as query data to assess model performance. The enumeration of drug-resistant and drug-sensitive cells within these datasets is presented in Fig. 2a. It is noted that the collected bulk-seq datasets consistently show a significant data imbalance issue, with the average number of drug-resistant cells (634.3) substantially exceeds that of drug-sensitive cells (82.4). To address this issue, we introduced several loss functions tailored for anomaly detection tasks to replace cross-entropy, shifting the focus from a standard binary classification of drug-sensitive versus resistant cells to identifying the sparse drug-sensitive cells among a majority of drug-resistant ones. These specialized loss functions improve accuracy by applying stronger penalties for misclassifying the less prevalent category, thus effectively prioritizing the minority class amid data imbalances. Fig. 2b illustrates the predictive performance of the five tested loss functions (Inverse(13), Deviation(14, 15), Hinge(16), Minus(17), Overlap(18)) on the Afatinib drug dataset across four performance metrics: Area Under the ROC Curve (AUROC), Area Under the Precision-Recall Curve (AUPR), Accuracy (ACC), and F1 Score (F1-macro). We observed significant improvements over the cross-entropy loss with average increases of 0.1002, 0.1019, 0.1450, and 0.1605 in AUROC, AUPR, ACC, and F1-macro, respectively. Notably, the scGSDR equipped with the Overlap(18) loss function (scGSDR-overlap) demonstrated the best predictive performance, achieving average values of 0.9219 in AUROC, 0.9352 in AUPR, 0.8583 in ACC, and 0.8567 in F1-macro.

We compared our models against two other methods specifically designed for cellular drug response annotation, SCAD and scDEAL, using the same validation framework. In Fig. 2c, the violin plot is constructed from ten points representing the results of ten experiments, and the red line on the plot depicts the mean values. The results illustrate the superior performance of our models, particularly the scGSDR-overlap model, across all metrics. Specifically, scGSDR-overlap significantly outperforms not only our baseline model using cross-entropy (scGSDR-ce) but also SCAD and scDEAL. In AUROC, the mean score for scGSDR-overlap is 0.8758, markedly higher than scGSDR-ce's 0.7676, and substantially outpaces SCAD's 0.6085 and scDEAL's 0.4647. Similarly, in AUPR, scGSDR-overlap achieves a mean score of 0.8861 compared to scGSDR-ce's 0.8001, 0.6753 for SCAD, and 0.5363 for scDEAL. Accuracy and F1-macro metrics exhibit similar patterns, with scGSDR-overlap reaching 0.8381 and 0.8263 in mean scores, respectively, demonstrating its effectiveness in predicting more accurately and maintaining a balance between precision and recall compared to other methods. The median values further confirm the robustness of scGSDR-overlap, showcasing its consistency in delivering higher performance.

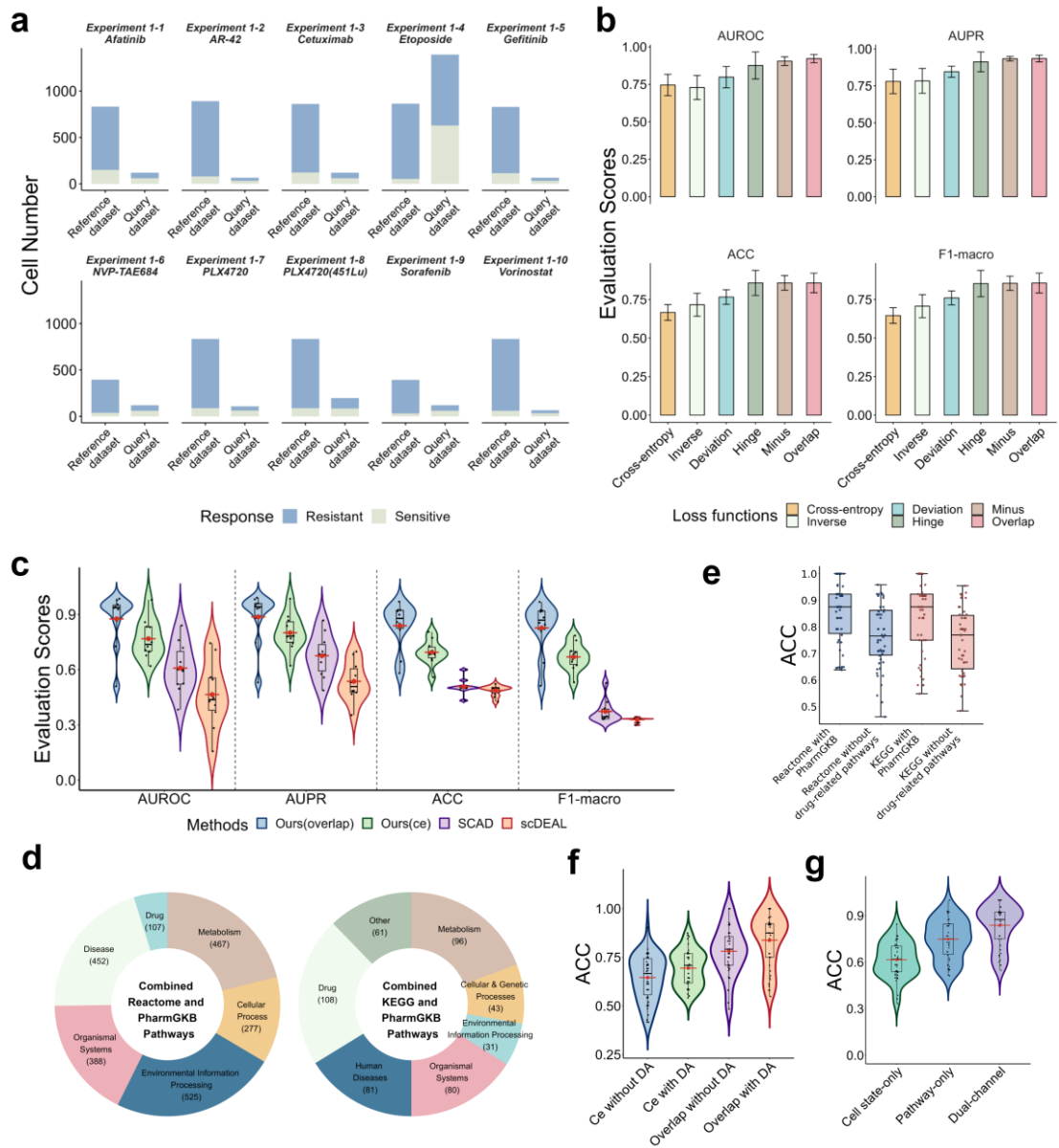

**Fig. 2** Dataset description and performance evaluation for scGSDR in experiments using bulk-seq reference datasets. **a** Datasets statistics. **b** Comparison of the model's performance under different loss functions across the metrics AUROC, AUPR, ACC, and F1-macro. **c** Performance comparison of scGSDR with benchmark methods. **d** Details of used pathways organized according to subsets of biological functions, with additional drug pathways sourced from PharmGKB supplementing the KEGG pathway database and Reactome pathway database. **e** Comparison of model performance regarding the use of drug pathways under the ACC metric. **f** Performance comparison regarding the use of domain adaptation. **g** Performance comparison of the model when using single and dual computational pipelines.

In the scGSDR model, genes are grouped according to a list of biomarkers related to cellular states and databases of gene signaling pathways. Their expressions across cells are funneled into two computational pipelines designed to facilitate the representation learning of cells. In the computational pipeline for gene signaling pathways, we group various pathways based on their attributes and use an attention mechanism to determine the contribution of each pathway to the final predictions of drug responses. To fully consider the interactions of genes within these pathways, we have gathered comprehensive gene signaling pathway data from the public databases Reactome(19) and KEGG(20). Considering the limited number of drug action pathways available in these databases (three in Reactome and four in KEGG), we have additionally collected 104 drug action pathways from PharmGKB(21). These have been merged with the existing data from Reactome and KEGG to create two enriched pathway datasets. Fig. 2d displays the statistical distribution of classifications based on attributes (Metabolism, Cellular Processes, Environmental Information Processing, Organismal Systems, Human Diseases, Drug Action) within the two pathway datasets. To assess the effectiveness of drug action pathways, we compared the predictive performance of scGSDR with and without the integration of drug action pathways across ten experimental conditions. The results (Fig. 2e) indicate that the removal of drug action pathways led to a decline in performance across both databases. Specifically, the average accuracy decreased from 0.8475 to 0.7636 in the Reactome experiments and from 0.8381 to 0.7462 in the KEGG experiments. To further test the efficacy of the DA classifier in the scGSDR model for addressing batch effects between reference and query datasets, we conducted a series of cross-experimental comparisons combining cross-entropy loss and overlap loss to assess the impact of adding DA. Fig. 2f presents a box plot illustrating the overall distribution and median levels of the result data, using ACC as the performance metric, with the mean indicated by a red line. The results demonstrate that the use of cross-entropy loss and overlap loss with DA led to average increases in prediction accuracy of 0.0484 and 0.0568, respectively. This underscores the importance of the DA classifier in mitigating batch effects across datasets. Fig. 2g illustrates a comparison between using a single computational pipeline and a dual-channel approach in the scGSDR model. The ablation study highlights the importance of each channel in the dual-channel model and the performance changes when a module is removed. When scGSDR uses only the cell state pipeline, it achieves an average accuracy of 0.6159 across ten experimental groups. Conversely, using solely the gene signaling pathway pipeline results in an average accuracy of 0.748. However, when both pipelines are employed simultaneously, the model reaches an average accuracy of 0.8381.

**2.3 Evaluation of scGSDR on scRNA-seq Reference Datasets**

As scRNA-seq technique gains prominence in drug research, it becomes feasible to use scRNA-seq data instead of cell line bulk-seq data for training models to predict drug responses at the single-cell level. As an extension of experiments based on bulk-seq data model training, we here further tested the performance of the scGSDR model when using scRNA-seq data as the reference data. We collected seven scRNA-seq datasets from three different drug treatment regimens to organize four experimental groups (Fig. 3a). The first and second experiments were based on the single-drug PLX4720. In Experiment 2-1, both reference and query datasets came from the 451Lu cell line, aiming to test the scGSDR model's performance in cross-platform scenarios and its ability to use small datasets for training to predict larger datasets. Experiment 2-2 used 451Lu as the reference and A375 as the query to evaluate the scGSDR model's predictive ability across different cell lines. Experiment 2-3 involved data collected from patients treated with a combination of Paclitaxel and Atezolizumab, assessing the scGSDR model's efficacy in multi-drug treatment scenarios. Experiment 2-4 utilized data from patients treated with Paclitaxel alone to test the scGSDR model's performance across tissue types in single-drug treatment experiments.

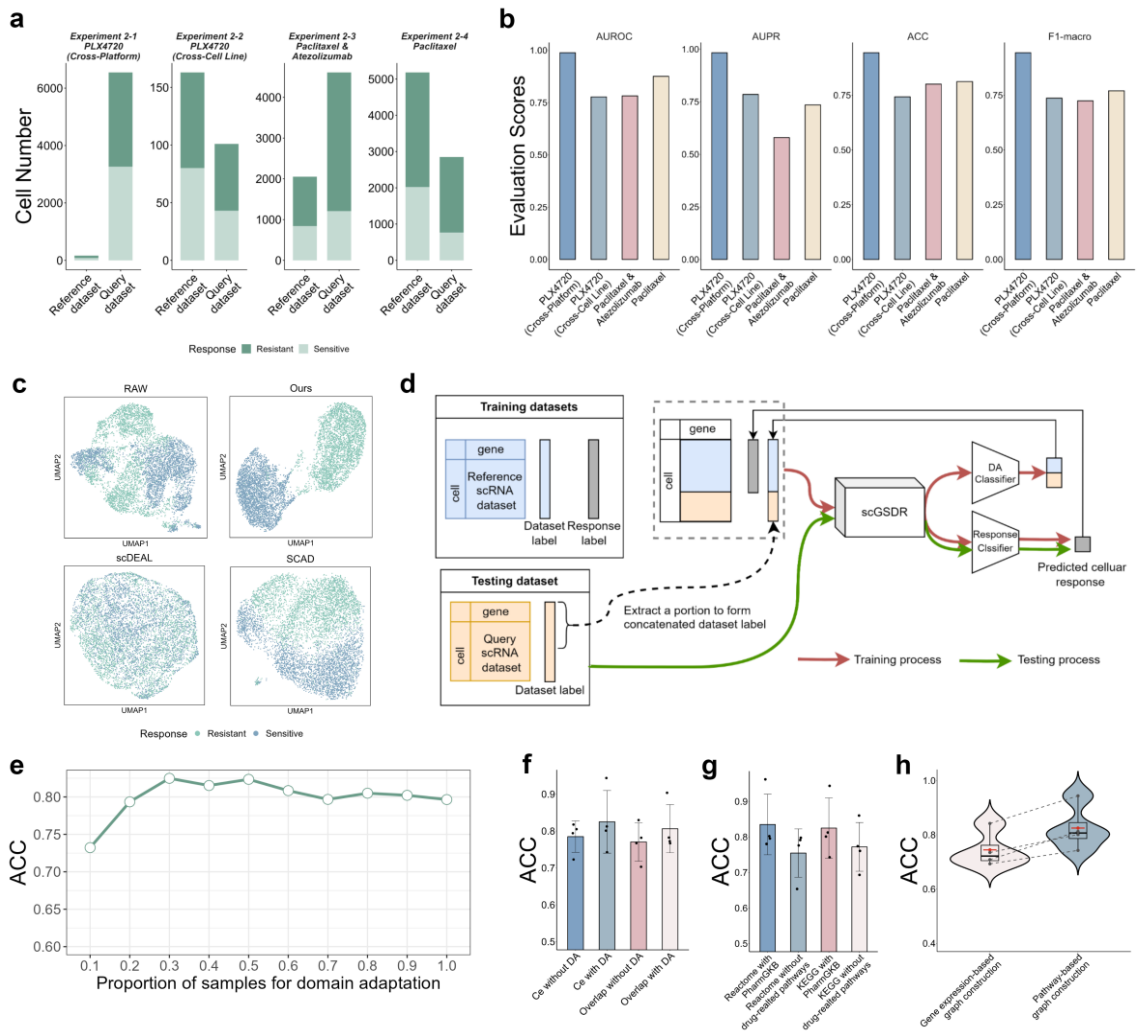

**Fig. 3** Dataset description and performance evaluation for scGSDR in experiments using scRNA-seq reference datasets. **a** Dataset statistics. **b** Comparison of the model's performance across various experiments. **c** UMAP projections of cells to compare the cell drug response prediction capabilities of scGSDR against other benchmark methods using data from Experiment 2-1. **d** Overview of domain adaptation training using source and target domain data in the model. **e** Comparison of model performance at different sampling ratios used for domain adaptation. **f** Performance comparison regarding the use of domain adaptation. **g** Performance Comparison regarding the use of drug pathways. **h** Performance Comparison between pathway-based and gene expression-based graph constructions.

The results in Fig. 3b show scGSDR's predictive performance across four experiments using four metrics. It is demonstrated that scGSDR achieved an average accuracy of 0.825 with a standard deviation of 0.0735 across all experiments. The model reached its highest accuracy of 0.944 in the first experiment and the lowest at 0.743 in the second, while the third and fourth experiments registered accuracies of 0.801 and 0.812, respectively. These results highlight the model's robust predictive capabilities, particularly in the cross-platform experiment, and its consistent performance across varied scenarios. In Fig. 3c, we further present the distribution of representations for each cell derived from the scGSDR, scDEAL, and SCAD models in the first experiment. It is shown that our scGSDR model is capable of effectively distinguishing between drug-resistant and drug-sensitive cells within the representation space, compared to raw data, and also outperforms the scDEAL and SCAD models. To mitigate batch effects between the reference and query datasets, we incorporated DA into the training of scGSDR, as depicted in Fig. 3d. A portion of the testing dataset (without cell type labels) was used to train the DA classifier. We evaluated the impact of using different proportions of the testing data on the average accuracy of scGSDR across four experiments. The results (Fig. 3e) indicate that introducing 30% of the testing samples was sufficient for scGSDR to eliminate the heterogeneity between datasets while maintaining stability upon the introduction of additional testing samples. The results (Fig. 3f) indicate that without using DA, the average accuracy of scGSDR trained with cross entropy and Overlap respectively decreased from 0.825 to 0.785 and from 0.806 to 0.770 across four experiments. To assess the utility of drug action pathways, we compared the predictive performance of scGSDR with and without the incorporation of PharmGKB pathways across ten experiments. The results (Fig. 3g) show that the removal of drug action pathways led to a decline in performance across both databases. Specifically, the average accuracy decreased from 0.835 to 0.755 in the Reactome experiments, and from 0.825 to 0.772 in the KEGG experiments. This indicates that the prior information provided by drug action pathways is beneficial for distinguishing cell responses to drugs. Fig. 3h presents a comparison of scGSDR's performance in constructing cell-cell graphs using both pathways and solely gene expression information across four experiments. The results demonstrate that removing the prior knowledge from gene pathways causes a decrease in the average accuracy of scGSDR from 0.825 to 0.745. This underscores the utility of the inherent gene relationships within pathways in enhancing the construction of cell-cell graphs. We further compare with 12 existing methods including machine learning-based cell type annotation tools such as SciBet(22), scPred(23), and singlecellNet(24); statistical methods like CHETAH(25); reference dataset-based tools such as SingleR(26); deep generative models including scVI(27) and scANVI(27); and marker gene-based tools like MarkerCount(28), each offering distinct approaches for achieve cell type classification in scRNA-Seq data analysis. As shown in Table 1, our method achieved best performance across three experiments and exhibited the highest average accuracy in four experiments. Notably, except for CHETAH, the comparison methods demonstrated instability in the four experiments, with standard deviations as high as 0.253. In contrast, the performance of our scGSDR model was comparatively stable, underscoring its robustness.

Table 1. Performance comparison of scGSDR with other benchmark methods for cell-type annotation under the ACC metric (Best performances are highlighted in bold, second best are underlined).

|  | Experiment 2-1 | Experiment 2-2 | Experiment 2-3 | Experiment 2-4 | Average |
|---|---|---|---|---|---|
| sciBet(22) | 0.958 | 0.465 | 0.744 | 0.816 | 0.746±0.207 |
| scPred(23) | 0.498 | 0.376 | 0.733 | 0.799 | 0.602±0.198 |
| CHETAH(25) | 0.796 | 0.723 | 0.734 | 0.812 | 0.766±0.044 |
| Seurat(29) | 0.498 | 0.079 | 0.746 | 0.804 | 0.532±0.330 |
| SingleR(26) | 0.955 | 0.455 | 0.748 | 0.744 | 0.726±0.205 |
| MarkerCount(28) | 0.876 | 0.158 | 0.734 | 0.614 | 0.595±0.311 |
| singlecellNet(24) | 0.956 | 0.436 | 0.769 | 0.783 | 0.736±0.217 |
| scANVI(27) | 0.954 | 0.455 | 0.753 | 0.773 | 0.734±0.207 |
| scPoli(30) | 0.497 | 0.495 | 0.749 | **0.832** | 0.643±0.173 |
| scLearn(31) | 0.788 | 0.089 | 0.653 | 0.074 | 0.401±0.373 |
| scmapCell(32) | 0.943 | 0.149 | 0.699 | 0.626 | 0.604±0.332 |
| scClassify(33) | 0.954 | 0.396 | 0.726 | 0.692 | 0.692±0.229 |
| Ours | **0.962** | **0.802** | **0.78** | 0.796 | **0.835±0.085** |

**2.4 Application of scGSDR on Combination Drug Therapy**

We further evaluated the predictive performance of the scGSDR model in the context of combination drug therapy experiments. Specifically, we investigated whether it is feasible to use datasets from individual drug treatments as references to train the scGSDR model. By leveraging this model, we aimed to develop a representation space that effectively distinguishes between drug-resistant and drug-sensitive cells in scenarios where the two drugs are used in combination. We here designed two experiments to further assess the scGSDR model (Fig. 4a): Experiment 3-1 involved the combination therapy of JQ1 and Paclitaxel, while Experiment 3-2 focused on the combination of JQ1 and Palbociclib. Fig. 4b shows the training framework for our combination drug prediction model. We augmented two monotherapy reference datasets with dataset labels and trained the scGSDR using a shared dataset classifier. This strategy aims to enable the scGSDR model to effectively predict the cellular drug response types common to both datasets, while also distinguishing and preserving the unique characteristics of each drug dataset. We utilized data from combined drug treatments as a test set to evaluate whether the scGSDR model can distinguish between drug-resistant and drug-sensitive cells within this dataset. To assess the performance, we employed clustering evaluation metrics based on the embedding space generated by the scGSDR's response classifier. We applied two unsupervised clustering algorithms, K-means and Birch, for cell clustering. The results (Fig. 4c) show that using K-means with scGSDR cellular embeddings significantly outperformed raw data across three metrics: Adjusted Rand Index (ARI), Normalized Mutual Information (NMI), and Average Silhouette Width (ASW). In Experiment 3-1, the improvements were 0.884, 0.708, and 0.161, respectively, while in Experiment 3-2, they were 0.651, 0.593, and 0.307, respectively. Similarly, with Birch, the enhancements in Experiment 3-1 were 0.654, 0.585, and 0.233, and in Experiment 3-2, they were 0.540, 0.514, and 0.311, respectively. Fig. 4d illustrates that scGSDR effectively distinguishes between drug-resistant and drug-sensitive cells, demonstrating improved intra-class cohesion and inter-class dispersion compared to raw data across two experiments. Our experiments demonstrate that scGSDR can effectively use data from individual drug treatments to accurately annotate cellular drug responses in scenarios involving combined drug use, without requiring experimental data from the combined administration of both drugs.

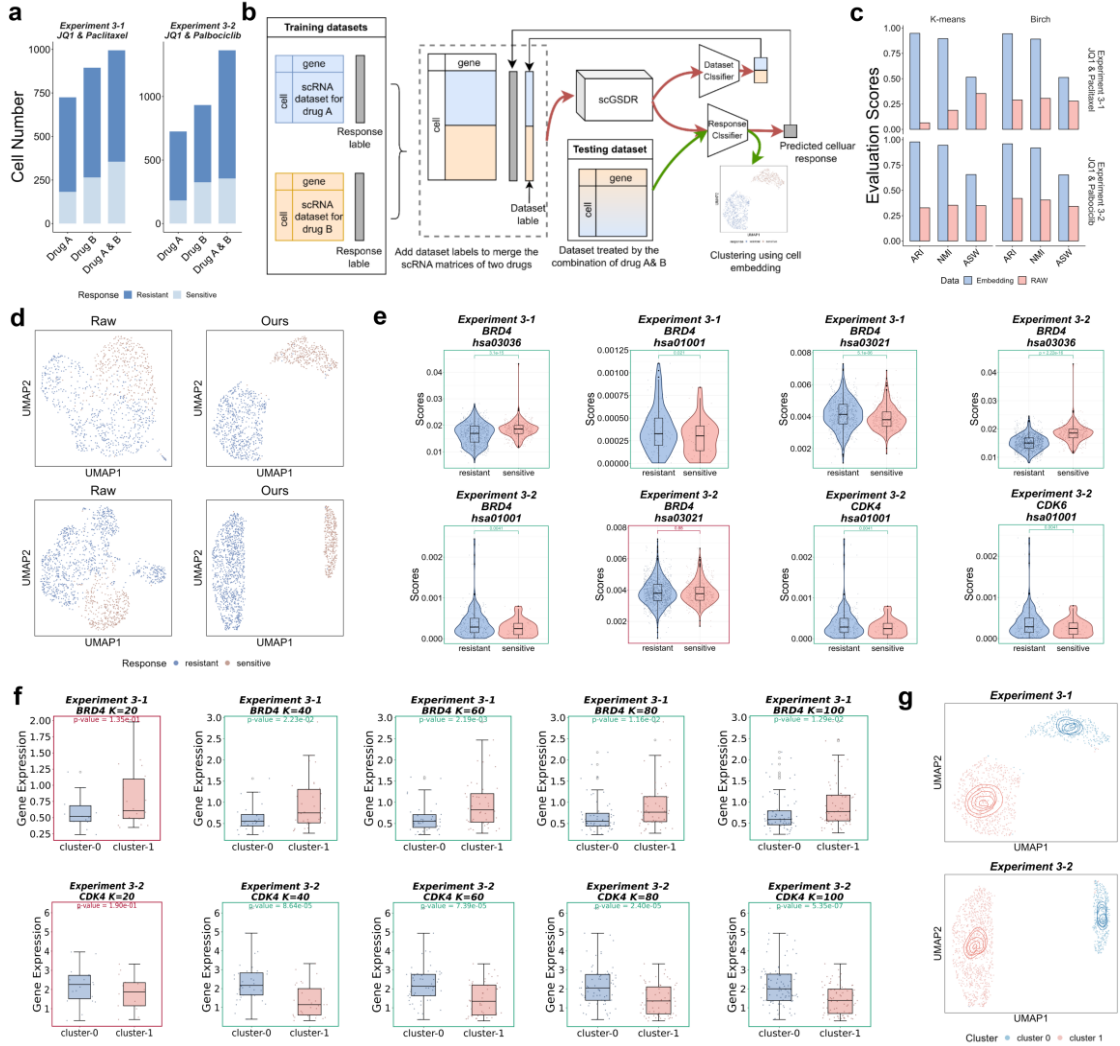

**Fig. 4** Dataset description and result analysis for scGSDR in drug combination experiments. **a** Dataset statistics. **b** Overview of training and testing strategies in combination drug datasets. **c** Comparison of clustering results under ARI, NMI, ASW metrics, using K-means and Birch clustering algorithms. **d** UMAP projection of cell using raw data and embeddings yielded by different methods. **e** Distributions of attention scores learned from scGSDR in two groups of drug-resistant and drug-sensitive cells (green indicates p-value < 0.05, red indicates otherwise). **f** Distribution of expression levels for known drug target genes (BRD4 and CDK4) between two groups, each consisting of the nearest K cells around two cluster centers (K=20, 40, 60, 80, 100); the upper half corresponds to Experiment 3-1, and the lower half to Experiment 3-2. **g** Visualization of the nearest K cells in two predicted cell clusters.

## 2.5 Analysis of Cell-pathway Attention in scGSDR with Biological Interpretability

In the pathway pipeline of scGSDR, each cell automatically learns the importance of all pathways through an attention mechanism, and these importance levels serve as weights that influence the drug response predictions of scGSDR. We evaluated the biological interpretability of these cell-pathway attention scores. Specifically, we first focused on the known target genes of drugs used in combination therapy experiments, observing whether the pathways containing these target genes received significantly different attention scores between drug-resistant and drug-sensitive cell groups. We examined gene pathways associated with the target gene BRD4 of drug JQ1 (hsa03036, hsa01001, and hsa03021) as well as the pathways associated with target genes CDK4 and CDK6 of drug Palbociclib (hsa1001). We collected attention scores from drug-resistant and drug-sensitive cell groups for these pathways and calculated the p-values between groups. The results (Fig. 4e) illustrate the distribution of these pathways across the two cell groups, with green indicating a p-value <0.05 and red otherwise. Except for hsa03021 in Experiment 3-2, all other results demonstrated significantly different attention scores between drug-resistant and drug-sensitive cells learned via scGSDR, suggesting that the attention mechanism offers substantial biological interpretability. We further tested the central cell collections in the clustering space for two types of drug response cells, specifically analyzing their expression levels of target drug genes (BRD4 for JQ1 and CDK4 for Palbociclib). As depicted in Fig. 4g, starting from the two cluster centers, we segmented the cells into five groups ranging from 20 to 100 cells at intervals of 20. The results (Fig. 4f) show that, except for the smallest group (20 cells), there were statistically significant differences in expression levels between the drug-resistant and drug-sensitive cell collections across all other groups. The larger the number of cells, the more statistically significant the differences, affirming the biological interpretability of the scGSDR model's learned cellular representation space and the clusters' representativeness for the two types of drug responses.

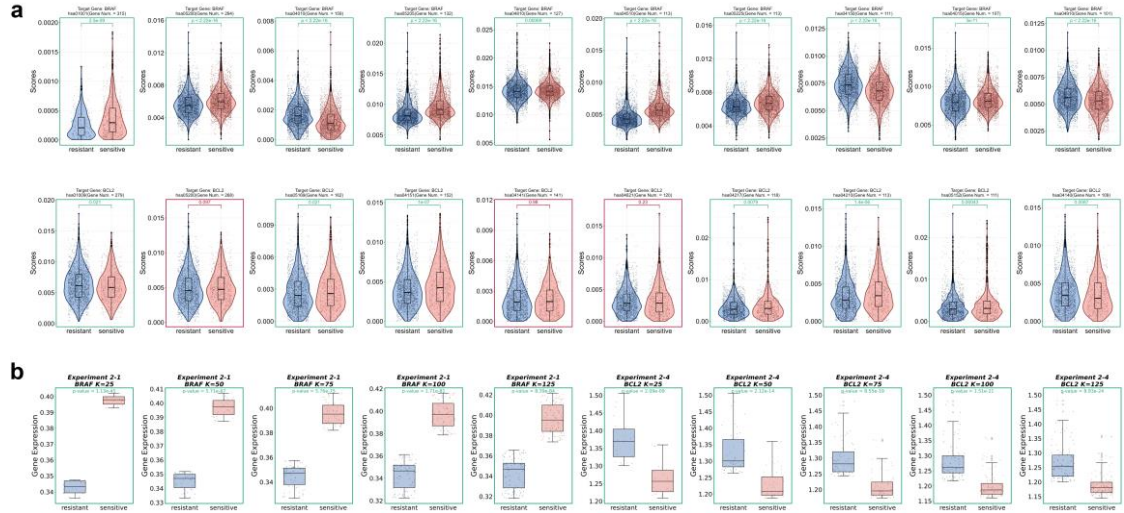

**Fig. 5** Statistical analysis of pathways and genes related to drug responses in experiments using scRNA-seq reference datasets. **a** Distribution of cell-pathway attention scores learned from scGSDR for the top 10 pathways with the highest number of genes (green indicates p-value < 0.05, red indicates otherwise). The upper half is from Experiment 2-1, and the lower half is from Experiment 2-4. **b** Distribution of expression levels for known drug target genes (BRAF and BLC2) between two groups, each consisting of the nearest K cells around two cluster centers (K=25, 50, 75, 100, 125). The first five bar scatter plots on the left are from Experiment 2-1, and the subsequent five are from Experiment 2-4.

Next, we similarly analyzed the cell-pathway attention in scGSDR for the two drugs, PLX4720 and Paclitaxel (experiments 2-1 and 2-4), to assess its biological interpretability in single-drug scenarios. We focused on the target genes BRAF and BCL2 for the drugs PLX4720 and Paclitaxel, respectively. We evaluated the attention scores for the top ten pathways involving the target gene with the highest number of genes, out of 43 pathways for BRAF and 37 for BCL2, in both drug-resistant and drug-sensitive cell groups. We calculated the p-values using t-tests to assess the statistical significance of the differences in attention scores between these groups. The results (Fig. 5a) show that in the PLX4720 experiment (Experiment 2-1), the attention scores for the top-10 pathways by gene count showed statistically significant differences between the two cell groups. In the Paclitaxel experiment (Experiment 2-4), except for the second, fifth, and sixth pathways (hsa05200, hsa04141, hsa04621), the remaining pathways also demonstrated statistically significant differences in attention scores between the two groups. In the cell embedding space learned from the scGSDR model, we centered on the cluster centroids of two types of cells. For each centroid, we selected its nearest K neighbors (K = 25, 50, 75, 100, 125) to form groups. We then calculated the statistical significance of the expression levels of drug target genes between the two cell types within each group. The results (Fig. 5b) showed significant differences in target gene expression at the cluster centroids for all ten group comparisons involving the two drugs. This demonstrates that the cluster centroids within the latent space learned through scGSDR can reliably identify representative clusters of drug-resistant and drug-sensitive cells.

**2.6 Application of scGSDR on Discovering Potential Gene Targets with Intersecting Pathways**

Our analyses so far have expanded from known drug target genes to gene pathways that they involve. We now aim to determine if pathways with significant statistical differences can reveal potentially other unknown drug-related genes. Our hypothesis posits that genes potentially related to drugs can be enriched across multiple gene pathways. Given the biological interpretability of the cell-pathway attention scores calculated by scGSDR in previous analyses, we seek to identify pathways with significant statistical differences in attention scores between two predicted cell clusters (drug-resistant and drug-sensitive). By examining these pathways, we aim to identify common genes within them, which could help infer potential genes associated with drug response outcomes.

Based on Experiment 2-1, we used this method to further infer other genes associated with the drug PLX4720. Specifically, we first collected the cell-pathway attention scores, which were pre-calculated using the scGSDR model with the query dataset (GSE108394) serving as the testing set. We then computed the p-values for each pathway using a t-test between the drug-resistant and drug-sensitive cell groups. Fig. 6a illustrates the distribution of attention scores on the top 20 pathways with the lowest p-values, based on the 100 genes with the largest variance in gene expression. It is shown that within the same pathway, the scores are similar within both cell groups, yet there are notable differences between the two groups. Based on the pathways with the lowest p-values for attention scores between cell groups, we identified the most frequently occurring genes within these pathways. These genes are considered potential candidates most relevant to the cellular response to the drug. We further identified the genes that appear most frequently within these pathways. These genes are considered the most relevant potential candidates related to the cellular response to the drug. Fig. 6b displays the occurrence of each gene in the pathways: the top panel is based on statistics from the top 20 pathways with the lowest p-values, while the bottom panel is based on the top 10. In the enrichment results of the top-20 pathways, we found that the top-1 gene, CCND1, was enriched in eight pathways. The distribution of attention scores for these eight pathways across all cells is depicted in Fig. 6c. Furthermore, based on the enrichment results from the top-10 pathways, the gene BCL2, which is enriched in three pathways (hsa04340, hsa04215, hsa04210), emerges as the most potentially relevant gene associated with the drug PLX4720, as visualized in Fig. 6d.

We further validated the gene lists enriched by intersecting pathways in experiments 2-1 and 2-4 through a review of existing literature. In the gene list generated from the top-10 pathways, BCL2 (rank 1) was inferred to be associated with PLX4720, while ICAM1 and VCAM1 (both rank 1, enriched by pathways hsa04064, hsa04090, hsa04091) were most likely associated with Paclitaxel. For BCL2, (34) found that the regulation of its family proteins may be associated with resistance to PLX4720 treatment in melanoma, particularly in the context of AKT pathway activation. It is worth noting that the AKT family gene is also present in the top-20 pathways in our experiment (upper panel in Fig. 6b). For ICAM1 and VCAM1, (35) discovered that low-dose Paclitaxel induces the expression of these inflammatory adhesion molecules in primary tumor samples.

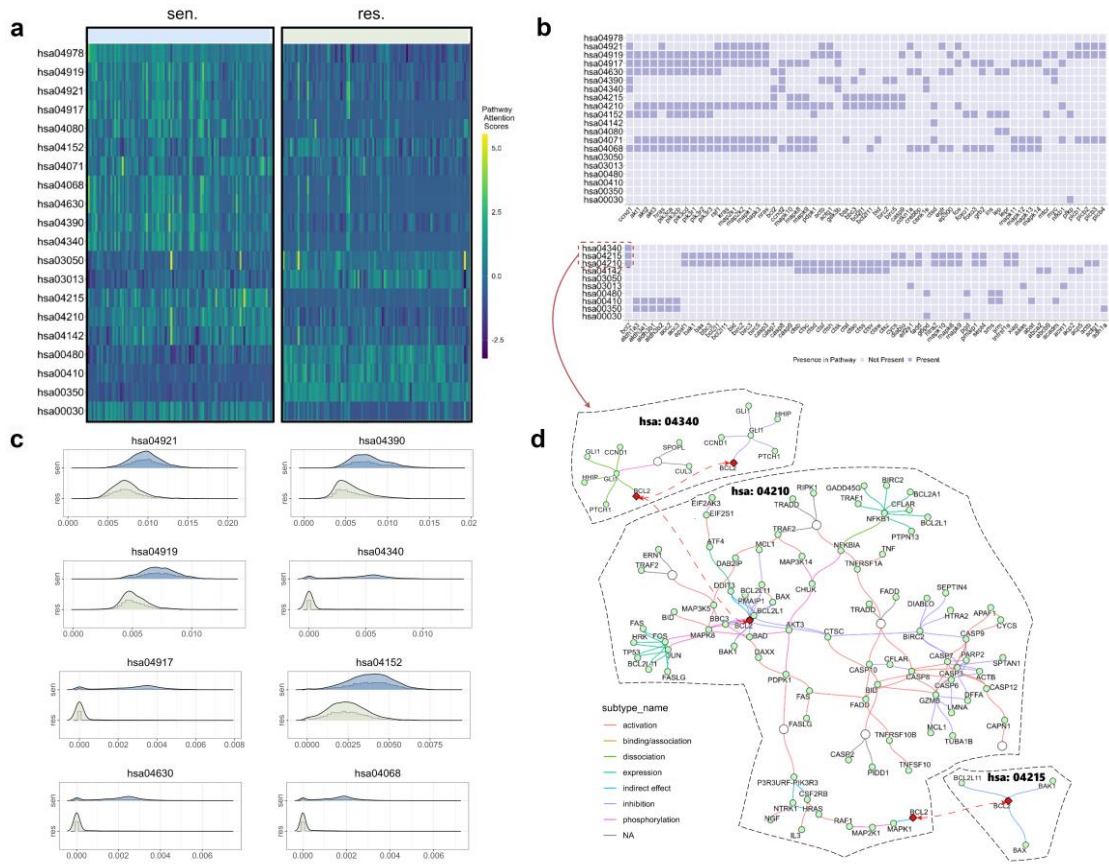

**Fig. 6** Application of cell-pathway attention learned from scGSDR in Experiment 2-1 to identify drug-associated genes that distinguish between resistant and sensitive cell phenotypes. **a** Heatmap distribution of attention scores for the top 20 pathways with the highest statistical significance (based on the smallest p-values from t-tests) in Experiment 2-1. The selection features 100 cells with the greatest variance in gene expression. **b** Display of shared genes within the collection of pathways with the highest statistical significance. The top portion features genes from the top 20 pathways, while the bottom portion includes those from the top 10 pathways, highlighting the top 60

genes most involved across these pathways. **c** Visualization of attention scores for eight pathways involving CCND1 (the rank-1 gene shared among the top-20 pathways), comparing their distribution across two different groups of cells. **d** Visualization of a gene pathway network involving BCL2 (the rank-1 gene shared among the top-10 pathways), constructed from partial structures of pathways hsa:04340, hsa:04210, and hsa:04215.

In the top-10 gene list enriched by the top-20 pathways for the drug PLX4720, we have identified through literature review that an additional five genes are associated with it: CCND1, AKT family(AKT1, AKT2, AKT3), and PIK3CA. Specifically, (36) found that PLX4720 may impact the expression or function of CCND1 in thyroid cancer cells by modulating signaling pathways influenced by the BRAFV600E mutation. (34) found that AKT activation inhibits BAD, enhancing cell survival and contributing to resistance to PLX4720. (37) found that CRC cell lines with PIK3CA mutations and/or PTEN loss were more resistant to growth inhibition by PLX4720 compared to cell lines without these alterations. Similarly, in the top-10 gene list enriched by the top-20 pathways for the drug Paclitaxel, three genes are confirmed: NFKB1, NFKBIA, RAC1. Specifically, (38) found that NFKB1 represses miR-134 expression, leading to TAB1 upregulation and contributing to paclitaxel resistance in ovarian cancer cells. (39) found that NFKBIA is significantly down-regulated by Paclitaxel, leading to NF-kB activation and contributing to Paclitaxel-induced apoptosis. (40) found that RAC1 promotes paclitaxel resistance in ovarian cancer cells by inhibiting pyroptosis through the PAK4/MAPK signaling pathway, thereby reducing the antitumor effect of paclitaxel.

## 3. Discussion

Predicting the response phenotype of individual cells to specific drugs based on differential gene expression profiles is essential, as it provides insights into the molecular mechanisms of drugs at the single-cell level, and is a critical step in applying scRNA-seq technology to pharmacological research and analysis. Existing computational methods for single-cell drug response prediction predominantly focus on utilizing bulk-seq cell line data as a reference for training models. In our work, we propose the scGSDR model, which extends to experimental scenarios involving single and combination drug treatments using scRNA-seq data as the reference. However, computational scenarios leveraging scRNA-seq as a reference pose new challenges, as the drop-out events inherent to scRNA-seq data induce sparsity in gene expression. This sparsity makes it increasingly difficult to discern the limited number of genes relevant to drug mechanisms, which are crucial for cell representation, from among a multitude of other genes. To address this issue, we propose the incorporation of prior knowledge of gene biological semantics to enhance the model's ability to learn cellular characterization. Considering that genes function collaboratively in groups through signaling pathways to perform certain biological functions, influencing the cellular state and ultimately determining the cell's response phenotype to a specific drug, we have designed two computational pathways within the scGSDR model that consider the biological semantic prior knowledge of genes in two aspects: (i) marker gene sets from different cellular states, and (ii) gene regulatory relationships within gene signaling pathways.

By incorporating prior knowledge of gene pathways, the scGSDR model implements a two-tiered coarse representation of genes. Initially, genes are clustered into a pathway-level representation based on the pathways they are involved in, using AUCell scoring. Subsequently, multiple pathways are aggregated according to biological function categories (as shown in Fig. 2d) and further represented at the biological function level through a cell-cell graph structure learner (as depicted in Fig. 1b). The introduction of such gene semantic information, as evidenced by the experimental results (Table 1), demonstrates that scGSDR can achieve both high predictive accuracy and good stability in experiments using scRNA-seq as a reference. Additionally, we validated across multiple datasets that the scGSDR model's computational pipeline for cellular state and gene pathway has complementary effects on predictive performance (Fig. 2g), confirming the synergistic value of these two types of prior knowledge. By analyzing the attention scores obtained from the pathways of known drug target genes, we observed significant statistical differences in these scores across cell clusters responding to two types of drugs, demonstrating that the attention generated by scGSDR possesses biological interpretability.

The incorporation of gene semantics not only enhances the predictive performance of the scGSDR model but also, through an attention mechanism interpretability module, enables it to identify which gene pathways contribute most significantly to its predictions. This facilitates the identification of key gene pathways that may determine cellular drug response phenotypes. Compared to single genes, gene pathways are coarser-grained and have known biological functions, making the cell-pathway attention scores generated by scGSDR inherently more interpretable and robust when elucidating the mechanisms behind drug responses. We leveraged the interpretability at the pathway level to identify potential genes associated with drug responses, utilizing single-drug scRNA-seq data from experiments 2-1 and 2-4. Several genes (BCL2, CCND1, AKT family, PIK3CA for PLX4720; ICAM1, VCAM1, NFKB1, NFKBIA, RAC1 for Paclitaxel) were identified and confirmed through literature review, validating the biological interpretability of the scGSDR model's attention mechanism. It should be noted that this method for discovering potential genes does not require any prior information about drug target genes for inference. Furthermore, it allows for the analysis of the reasoning behind these inferences (i.e., which gene pathways are relevant to the inferred genes), thereby providing valuable application in the study of potential drug mechanisms.

The current version of scGSDR still has some limitations, which we hope to address and improve in future work. The training of the scGSDR model relies on reference datasets and gene semantic data, meaning that selecting different datasets as references could lead to variations in performance and interpretability for the same query dataset. However, the availability of scRNA-seq experimental data for the same drug remains scarce, with only one set available for PLX4720 (experiment 2-1) and one for Paclitaxel (experiment 2-4). The uncertainty introduced by the choice of reference data still needs to be tested in the future. Moreover, since scGSDR utilizes gene pathways to group genes performing similar functions, genes involved in multiple pathways may disproportionately influence the final predictive outcomes, leading to potential oversight of critical genes that are involved in fewer pathways. Additionally, scGSDR still faces issues with data scarcity. If more than one set of data becomes available for reference in the future, scGSDR could potentially be iteratively trained across multiple reference datasets to achieve more robust and accurate performance.

## 4. Conclusions

scGSDR leverages gene semantic knowledge to enhance the accuracy of single-cell drug response predictions and provides interpretable insights into drug molecular mechanisms at the gene pathway level, significantly advancing the analysis of scRNA-seq data in drug resistance studies. Comprehensive experiments across various settings have confirmed that scGSDR delivers high performance on different types of training data, including both bulk-seq and scRNA-seq data, and is applicable to predictions for either single-drug or multi-drug combination trials. scGSDR not only achieves accurate and robust predictive performance across various experimental datasets but also provides the rationale for its predictions through an interpretability module, thereby identifying trustworthy gene pathways that have a significant impact on cellular drug response profiling. Based on these pathways, we examined the top-ranking list and successfully confirmed through literature review the identification of some potential drug-related genes, such as the BCL2 gene for PLX4720 and the ICAM1 gene for Paclitaxel, demonstrating scGSDR's potential application in identifying key genes involved in drug mechanisms. We anticipate that scGSDR will serve as a valuable tool, providing accurate cellular drug response profiling and facilitating the discovery of potential gene pathways in the analysis of single-cell drug response experiments.

## 5. Methods

**Datasets used in the experiments on bulk RNA-seq reference datasets:** In our study, 10 experiments (denoted as Experiment 1-1 to Experiment 1-10) were designed for predicting drug responses in scRNA-seq datasets using bulk RNA-seq datasets, with these experiments involving 9 different drugs. In these experiments, all reference datasets were obtained from the high-throughput pharmacogenomic database GDSC(10), consisting of transcriptome profiles and corresponding drug response sensitivity data. The query datasets for Experiments 1-1, 1-2, 1-3, 1-5, 1-6, 1-9, and 1-10 were derived from SCP542(41), while the query dataset for Experiment 1-4 was derived from GSE149215(42), and those for Experiments 1-7 and 1-8 were derived from GSE108383. Experiment 1-1 is related to the drug Afatinib, where the reference dataset consists of 832 cells (682 resistant cells and 150 sensitive cells), and the query dataset consists of 120 cells (60 resistant cells and 60 sensitive cells). Experiment 1-2 is related to the drug AR-42, where the reference dataset consists of 891 cells (811 resistant cells and 80 sensitive cells), and the query dataset consists of 66 cells (33 resistant cells and 33 sensitive cells). Experiment 1-3 is related to the drug Cetuximab, where the reference dataset consists of 861 cells (739 resistant cells and 122 sensitive cells), and the query dataset consists of 120 cells (60 resistant cells and 60 sensitive cells). Experiment 1-4 is related to the drug Etoposide, where the reference dataset consists of 864 cells (811 resistant cells and 53 sensitive cells), and the query dataset consists of 1393 cells (764 resistant cells and 629 sensitive cells). Experiment 1-5 is related to the drug Gefitinib, where the reference dataset consists of 829 cells (714 resistant cells and 115 sensitive cells), and the query dataset consists of 66 cells (33 resistant cells and 33 sensitive cells). Experiment 1-6 is related to the drug NVP-TAE684, where the reference dataset consists of 395 cells (358 resistant cells and 37 sensitive cells), and the query dataset consists of 120 cells (60 resistant cells and 60 sensitive cells). Experiment 1-7 is related to the drug PLX4720, where the reference dataset consists of 834 cells (746 resistant cells and 88 sensitive cells), and the query dataset consists of 108 cells (46 resistant cells and 62 sensitive cells). Experiment 1-8 is related to the drug PLX4720, where the reference dataset consists of 834 cells (746 resistant cells and 88 sensitive cells), and the query dataset consists of 197 cells (113 resistant cells and 84 sensitive cells). Experiment 1-9 is related to the drug Sorafenib, where the reference dataset consists of 393 cells (362 resistant cells and 31 sensitive cells), and the query dataset consists of 120 cells (60 resistant cells and 60 sensitive cells). Experiment 1-10 is related to the drug Vorinostat, where the reference dataset consists of 834 cells (774 resistant cells and 60 sensitive cells), and the query dataset consists of 66 cells (33 resistant cells and 33 sensitive cells). The data for Experiments 1-1 to 1-10 are available from SCAD.

**Datasets used in the experiments on scRNA-seq reference datasets:** As part of our study on predicting drug responses in scRNA-seq datasets using scRNA-seq datasets, 4 experiments were designed, denoted as Experiment 2-1 to Experiment 2-4. Experiment 2-1 is a cross-platform experiment involving the drug PLX4720, where the reference dataset is from GSE108383(43) and the query dataset is from GSE108394(43), using data generated on NextSeq and HiSeq platforms, respectively. The reference dataset consists of 163 cells (83 resistant cells and 80 sensitive cells), and the query dataset consists of 6548 cells (3262 resistant cells and 3286 sensitive cells). Experiment 2-2 is a cross-cell line prediction experiment involving the drug PLX4720, using data from GSE108383, where the reference dataset is from the 451Lu cell line and the query dataset is from the A375 cell line. The reference dataset consists of 163 cells (83 resistant cells and 80 sensitive cells), and the query dataset consists of 101 cells (43 resistant cells and 58 sensitive cells). Experiment 2-3 and 2-4 are cross-tissue prediction experiments, using patient PBMC data to predict tumor tissue cell drug responses, with data from GSE169246(44). Experiment 2-3 uses the drug combination Paclitaxel & Atezolizumab, with a reference dataset consisting of 2053 cells (1213 resistant cells and 840 sensitive cells), and a query dataset consisting of 4612 cells (3404 resistant cells and 1208 sensitive cells). Experiment 2-4 uses the drug Paclitaxel, with a reference dataset consisting of 5183 cells (3161 resistant cells and 2022 sensitive cells), and a query dataset consisting of 2851 cells (2086 resistant cells and 765 sensitive cells). The data for Experiments 2-1 to 2-4 are available from DRMref(45).

**Datasets used in the combination therapy experiments:** In the combination therapy experiments, expression data and drug responses under two single-drug treatments were used to train the model, and cell drug responses under the combination of the two drugs were predicted. The data used in the experiments came from GSE131984(46). Two experiments were designed, denoted as Experiment 3-1 and Experiment 3-2. In Experiment 3-1, the two drugs used were JQ1 and Paclitaxel. The JQ1 dataset consisted of 725 cells (543 resistant cells and 182 sensitive cells), the Paclitaxel dataset consisted of 895 cells (630 resistant cells and 265 sensitive cells), and the JQ1 & Paclitaxel combination therapy dataset consisted of 995 cells (640 resistant cells and 355 sensitive cells). In Experiment 3-2, the two drugs used were JQ1 and Palbociclib. The JQ1 dataset consisted of 725 cells (543 resistant cells and 182 sensitive cells), the Palbociclib dataset consisted of 933 cells (608 resistant cells and 325 sensitive cells), and the JQ1 & Palbociclib combination therapy dataset consisted of 1364 cells (1009 resistant cells and 355 sensitive cells). The data for Experiments 3-1 and 3-2 are available from DRMref.

**The data preprocessing workflow for the experiments on bulk RNA-seq reference datasets:**
For the experiment using bulk RNA-seq data to predict scRNA-seq data, the bulk RNA-seq dataset underwent RMA normalization and was standardized to z-scores. The scRNA-seq dataset was processed using the Scanpy(47), which included a series of preprocessing steps. These included various data quality control (QC) steps such as cell QC and gene QC based on dataset characteristics like minimum gene count, minimum total count, minimum cell count, and maximum total count. Additionally, cells with mitochondrial counts exceeding 20% of the total molecular counts were removed. Subsequently, the dataset was normalized by total counts over all genes, log-transformed, and standardized to z-scores. The aforementioned data processing workflow has been completed by the data provider.

**The data preprocessing workflow for the experiments on scRNA-seq reference datasets:** For the experiment involving the prediction of scRNA-seq data drug responses using scRNA-seq data, the datasets were preprocessed using Scanpy. Quality control was conducted on the scRNA-seq dataset, including cell QC based on minimum and maximum gene counts and minimum and maximum total counts, and gene QC based on the minimum number of cells. The data was then normalized to a total molecular count of 1e5, followed by log normalization and z-score standardization.

**The data preprocessing workflow for the combination therapy experiments:** In the combination drug experiment, within the Experiment 3-1 (JQ1 & Paclitaxel), the JQ1 dataset, the Paclitaxel dataset, and the JQ1 & Paclitaxel dataset were preprocessed separately. Preprocessing included cell QC, gene QC, and normalization of total molecular counts across all genes for each cell. The normalized datasets for JQ1 and Paclitaxel were concatenated to form the training dataset, while the JQ1 and Paclitaxel combination treatment dataset was used for testing. Since the sensitive cell data was shared among the JQ1, Paclitaxel, and JQ1 and Paclitaxel combination treatment datasets, the sensitive cells were divided into three parts and allocated to the respective datasets. The preprocessing workflow for the Experiment 3-2(JQ1 & Palbociclib) was consistent with that of the Experiment 3-1.

**Workflow Overview of scGSDR:** The scGSDR model learned the correlation between gene expression and drug response from the reference dataset and predicted the drug response of cells in the query dataset. scGSDR consists of three core components, including two computational pipelines and an attention-based feature fusion module. Specifically, scGSDR considers genes in scRNA-seq datasets from two biological semantic perspectives, (i) cellular states and (ii) cellular signaling pathways, through the two computational pipelines, and utilizes (iii) the attention-based feature fusion module to fuse the cellular features obtained from the two pipelines. In the first pipeline, we used marker genes of different cellular states as the criteria to filter genes in the dataset and constructed cellular features. The cellular features were mapped to an embedding space using a transformer-based feature extractor. In the second pipeline, cellular graphs that integrated cellular signaling pathways were automatically learned using a cellular graph learner based on dot-product self-attention, and a GCN-based feature extractor was used to aggregate features of neighboring cells and extract the embedding representation of cells. The cellular embeddings obtained from the two pipelines were fused by an attention-based feature fusion module to obtain the final fused embedding, which was input into a fully connected classifier to predict drug response. To enable the feature extractor to capture invariant features of cells in the reference dataset and query dataset, we introduced domain adaptation (DA) to ensure that the cellular features learned from the two datasets were domain-invariant.

**Cellular State Feature Extractor in Cellular States Pipeline:** This module took the expression matrix filtered by the gene sets of 14 cellular states as input, and then used a transformer to map it to an embedding space, and yielded representations of the cells. Gene sets associated with 14 cellular states were extracted from CancerSEA(48), including Angiogenesis (73 genes), Apoptosis (66 genes), Cell Cycle (137 genes), Differentiation (201 genes), DNA damage (110 genes), DNA repair (119 genes), EMT (90 genes), Hypoxia (83 genes), Inflammation (112 genes), Invasion (97 genes), Metastasis (166 genes), Proliferation (88 genes), Quiescence (66 genes), and Stemness (166 genes).

The expression matrix was filtered using the gene sets corresponding to each cellular state, resulting in 14 expression matrices. These matrices represented the expression levels of all cells in different cellular states. Define $N$ as the number of cells and $N_s$ as the number of cellular states. A linear layer was utilized to map the feature vectors of different cellular states to a unified length $N_e$, which was set to 16, resulting in a matrix $H_s \in \mathbb{R}^{N \times N_s \times N_e}$. A transformer was then employed to process these data. The transformer's output was denoted as $H_s' \in \mathbb{R}^{N \times N_s \times N_e}$, and through the process of matrix concatenation and linear transformation, the final output $Z_s \in \mathbb{R}^{N \times N_d}$ was obtained.

**Cellular Graphs Learner in Cellular Signaling Pathways Pipeline:** This module, based on dot-product self-attention, utilized the cell-pathway matrix as input to automatically learn the associations between each cell and various signaling pathways, subsequently constructing cellular graphs. The expression status of each cell in specific pathways was assessed using the AUCell method, and cell-pathway score matrices were generated for each subset of the pathway database, which was categorized based on biological function. Subsequently, each matrix was used to construct the corresponding cellular graph. The number of graphs was represented by $G$. The number of pathways included in each pathway subset varies, and each matrix is expanded to the maximum length, which is the maximum number of pathways across all subsets, denoted as $d_m$. Specifically, for matrices that did not reach the length of $d_m$, the corresponding values in the missing columns were set to zero. Let $N$ denoted the number of cells. By concatenating $G$ cell-pathway score matrices, each of length $d_m$, the matrix $H \in \mathbb{R}^{N \times G \times d_m}$ was obtained. The cellular graph learner based on dot-product self-attention was defined as:

$$A_{1:G} = Sigmoid\left(\frac{QK^T}{\sqrt{d_k}}\right), \qquad (1)$$

where $d_k$ represented the length of the key matrix and was set to be equal to $d_m$, $Q = HW_Q$, and $K = HW_k$. $Q$ and $K$ represented the query and key matrices, where $Q, K \in \mathbb{R}^{N \times G \times d_m}$. $W_Q$ and $W_k$ were trainable weight parameter matrices, where $W_Q, Q_K \in \mathbb{R}^{d_w \times d_w}$. In the cellular graphs learner, two levels of attention information were defined: pathway-level ($d_w = G \times d_m$) and functional-level ($d_w = G$). Sparsification was then performed on the Q matrix:

$$q'_{i,j,g} = ReLU\left(q_{i,j,g} - Sigmoid(\theta)\right), \tag{2}$$

where θ was a learnable parameter. The corresponding query (Q), key (K), and attention (A) matrices were subsequently calculated. A sparsification operation was performed on the adjacency matrix $A_g$:

$$a'_{i,j,g} = ReLU\left(a_{i,j,g} - threshold_k \times \theta_{scale}\right), \tag{3}$$

where $threshold_k$ was defined as the value of the $k$-th highest node in the matrix, and $\theta_{scale}$ was a learnable scaling parameter.

**Feature Extractor in Cellular Signaling Pathways Pipeline:** This module took the learned cellular graphs and filtered gene expression profiles as input, and extracted cellular features related to each graph using GCN. The entire pathway database was functionally categorized into distinct subsets, from which the genes were extracted. The genes in the expression matrix were then filtered to produce the feature matrices $F_{1:G}$, each corresponding to a specific pathway subset.

A two-layer fully connected network was utilized to transform the node feature matrices $F_{1:G}$ into a unified dimension. Subsequently, all the remaining node feature matrices were concatenated to form a new matrix, denoted as $F'_{1:G}$. The adjacency matrices learned by the cellular graphs learner were represented as $A'_{1:G}$. To learn the cellular representations in graphs, we employed a $L$-layers GCN. The GCN at the $l$-th layer was defined as follows:

$$GCN^{(l)}(A'_g, F'_g) = \widehat{D}_g^{-\frac{1}{2}} \widehat{A}_g \widehat{D}_g^{-\frac{1}{2}} F'_g W^{(l)}, \tag{4}$$

where $W^{(l)}$ was a trainable weight matrix, $\widehat{A}_g = A_g + I_N$ represented the adjacency matrix with added self-loops, where $I_N$ denoted an identity matrix, and $\widehat{D}_g = diag(\widehat{A}_g 1_{1 \times N}^T)$ was the degree matrix. The output of this part of the model was $Z_g \in \mathbb{R}^{N \times N_d}$. By concatenating the $G$ matrices, $Z_P \in \mathbb{R}^{N \times G \times N_d}$ was obtained.

**Feature Fusion Module Based on Attention:** This module was used to fuse the cellular features obtained from the cellular states pipeline (denoted as $Z_S \in \mathbb{R}^{N \times 1 \times N_d}$) and the signaling pathways pipeline (denoted as $Z_P \in \mathbb{R}^{N \times G \times N_d}$), and to obtain the final fused features. $Z_S$ and $Z_P$ were concatenated into $X \in \mathbb{R}^{N \times (G+1) \times N_d}$, where $N$ denoted the number of cells in the dataset, $G + 1$ represented the number of features to be fused, encompassing the feature from the cellular states pipeline and $G$ features from the signaling pathways pipeline, and $N_d$ represented the number of feature dimensions.

Assuming $\omega$ was the weight matrix computed by the linear layer and applied to each group of features in $X$, the attention score for each group of features was defined as:

$$\alpha = softmax\left(\frac{X\omega}{\tau}\right), \tag{5}$$

where $\tau$ was the temperature parameter, utilized to control the smoothness of the softmax function. Next, weighted summation of the features was conducted:

$$Z = \sum_{i=1}^{G+1} (X_i \cdot \alpha_i), \tag{6}$$

where $Z$ exhibited a shape of $[N, N_d]$.

**Loss Function for Drug Response Classifier:** The cross-entropy loss function was defined to evaluate the accuracy of the predicted drug responses, using the formula:

$$Loss_{CE} = -\frac{1}{N} \sum_{i=1}^{N} \sum_{c=1}^{C} y_{i,c} \log g(\hat{p}_{i,c}), \tag{7}$$

where $y_{i,c}$ was an indicator function that equals 1 when the true class of the $i$-th sample was $c$,

and 0 otherwise. $\hat{p}_{i,c}$ denoted the probability predicted by the model that the $i$-th sample belonged to class $c$. $N$ was the total number of samples, and $C$ was the number of classes.

This loss function was directly applied to the binary labels of the cellular drug responses, where $y$ was the true binary label and $\hat{y}$ was the predicted binary label, thus quantifying the prediction error across all classes.

An anomaly detection loss function, Overlap, was introduced to replace the Cross Entropy loss function. To comprehensively evaluate the performance of the Overlap loss function in comparison to other anomaly detection functions, we also compared the Minus, Inverse, Hinge, and Deviation loss function. In the definitions of the above loss functions, $s_n$ represented the score of normal samples, and $s_a$ represented the score of abnormal samples.

The Minus loss function can be represented as:

$$Loss_{Minus} = |s_n| + max(0, BND - |s_a|), \qquad (8)$$

where the hyperparameter $BND$ was set to 5.

The Inverse loss function can be represented as:

$$Loss_{Inverse} = |s_n| + 1/|s_a|. \qquad (9)$$

The Hinge loss function can be represented as:

$$Loss_{Hinge} = max(0, M + s_n - s_a), \qquad (10)$$

where the hyperparameter $M$ was set to 5.

The Deviation loss function can be represented as:

$$Loss_{Deviation} = |s_n| + max(0, M - s_a), \qquad (11)$$

where the hyperparameter $M$ was set to 5, and the anomaly scores were normalized as Z-Scores.

For convenience, these losses were collectively referred to as $Loss_{pred}$.

**Loss Function for Domain Adaptation Classifier:** To learn domain-invariant features from both the source and target domains, a domain adaptation classifier $f_{DA}(\cdot)$ was trained to compel the feature extractor. This domain adaptation mechanism facilitated the reduction of the discrepancy between the distributions of the source and target domains, thereby enhancing the performance of the drug response predictor. Upon completion of the training process, the feature extractor was capable of extracting feature representations that the domain adaptation classifier cannot distinguish as originating from the source or target domain, i.e., domain-invariant features.

The domain adaptation classifier was defined as:

$$f_{DA}(x) = \sigma\left(W^3\left(ReLU\left(W^2(ReLU(W^1 x))\right)\right)\right), \tag{12}$$

where $W_1$, $W_2$, and $W_3$ were the weight matrices of the linear layers and $\sigma$ was the sigmoid function. Assuming $Z$ was the cellular representation extracted by the model's feature extractor. The loss on the domain adaptation classifier was defined as:

$$Loss_{DA} = Loss_{CE}(Y_d, f_{DA}(Z)), \tag{13}$$

where $Y_d$ was the true value of the cellular domain affiliation, and $f_{DA}(Z)$ was the predicted value indicating the cellular domain affiliation. The domain adaptation classifier aimed to minimize $Loss_{DA}$, while the rest of the model was trained to maximize the loss.

**Loss Function for Graph Reconstruction:** The graph reconstruction loss was defined as the Mean Squared Error (MSE) loss between the graph generated by the embeddings derived from the model and the graph utilized by GCN. The MSE loss function was formulated as:

$$Loss_{MSE} = \frac{1}{n}\sum_{i=1}^{n}(y_i - \hat{y}_i)^2, \tag{14}$$

where $y_i$ denoted the true value of the $i$-th sample, $\hat{y}_i$ represented the predicted value of the $i$-

th sample, and $n$ signified the total number of samples. Assuming $Z$ was the feature matrix output from the model, the reconstructed graph was defined as:

$$\hat{A} = ZZ^T. \qquad (15)$$

The reconstructed graph $\hat{A}$ was sparsified using the threshold $\tau$, which was defined as the $k$-th largest value in $\hat{A}$ and $k$ was set to 20:

$$\hat{A} = ReLU(\hat{A} - \tau). \qquad (16)$$

Given that the adjacency matrices utilized in the GCN were represented by $A_1, A_2, \ldots, A_G$, the reconstruction loss was defined as:

$$Loss_{recon} = Loss_{MSE}(\hat{A}, A), \qquad (17)$$

where $A = \sum_{i=1}^{G} A_i$.

**Overall Loss Function of the Model:** The model's overall loss function comprised three components: the drug response prediction loss, denoted as $Loss_{pred}$, the domain adaptation loss, denoted as $Loss_{DA}$, and the graph reconstruction loss, denoted as $Loss_{recon}$. The final loss function of the model was defined as follows:

$$Loss = Loss_{pred} - \lambda_{DA} Loss_{DA} + \lambda_{recon} Loss_{recon}, \qquad (18)$$

where $\lambda_{DA}$ and $\lambda_{recon}$ represented the weight parameters assigned to each component of the total loss and were set to 1e-5.

**Evaluation Metrics:**

The model's performance was evaluated using four metrics: AUROC, AUPR, Accuracy, and F1-macro score. AUROC measured the probability that the model predicted a higher score for a randomly selected positive sample than for a randomly selected negative sample. AUPR calculated the area under the precision-recall (PR) curve, with the baseline of the PR curve varying in different tasks according to the proportion of positive samples in the total sample size. Accuracy denoted the proportion of correctly predicted samples out of the total sample size. F1-macro score was the harmonic mean of the model's precision and recall, which can be calculated as:

$$F1 - macro = 2 * (p * r)/(p + r), \qquad (19)$$

where $p$ represented precision and $r$ represented recall.

**Comparison Methods:**

The comparison methods used in the study include CHETAH (version 1.18.0), scANVI (version 1.1.5), scPoli (version 0.6.1), MarkerCount (version 0.6.12), scLearn (version 1.0), Seurat (version 4.3.0), scmapCell (version 1.24.0), scPred (version 1.9.2), scClassify (version 1.5.1), sciBet (version 0.1.0), SingleR (version 2.4.1), singleCellNet (version 0.1.0), SCAD (version not available), and scDEAL (release 1.0).

**Availability of data and materials**

The model is implemented in Python (version 3.9). The source code is available on Github (https://github.com/ChrisOliver2345/scGSDR) under the MIT license for academic researchers. In the experiment of training models using bulk RNA-seq and predicting drug responses in scRNA-seq data, the reference data for experiments 1-1 to 1-10 were sourced from GDSC(10). The query datasets for experiments 1-1, 1-2, 1-3, 1-5, 1-6, 1-9, and 1-10 were obtained from SCP542(41). The query dataset for experiment 1-4 was derived from GSE149215(42), while the query datasets for experiments 1-7 and 1-8 were sourced from GSE108383(43). In the study utilizing scRNA-seq datasets to predict drug responses within scRNA-seq datasets, the data for experiment 2-1 were sourced from GSE108383(43) and GSE108394(43), the data for experiment 2-2 were obtained from GSE108383(43), and the data for experiments 2-3 and 2-4 were derived from GSE169246(44). In the combination therapy experiments, the data used in the experiments were obtained from GSE131984(46).

**Keywords**

drug resistance, single-cell sequencing data, gene semantics, model interpretability